\documentclass[prl,letterpaper,twocolumn,aps,footinbib,superscriptaddress,showpacs,babel]{revtex4-1}
\usepackage{amsmath, upgreek, amssymb, graphicx, subfigure, paralist, dsfont,transparent}
\usepackage[colorlinks, linkcolor=blue, citecolor=blue, urlcolor=blue, breaklinks=true]{hyperref}
\usepackage{xspace}

\newcommand{\ket}[1]{|#1\rangle}
\newcommand{\bra}[1]{\langle #1|}
\newcommand{\braket}[1]{\langle #1\rangle}

\newcommand{\C}{\mathrm{C}}
\renewcommand{\L}{\mathrm{L}}
\newcommand{\Nc}{N_\C} 
\newcommand{\Nl}{N_\L} 
\def\fm#1{\ifmmode #1 \else $#1$\fi}
\def\Al{\fm{\mathrm{Al}^{+}}\xspace}
\def\Ca{\fm{\mathrm{Ca}^{+}}\xspace}

\begin{document}

\title{Quantum Algorithmic Readout in Multi-Ion Clocks}

\author{M. Schulte}
\affiliation{Institute for Theoretical Physics and Institute for Gravitational Physics (Albert-Einstein-Institute), Leibniz University Hannover, Callinstrasse 38, 30167 Hannover, Germany}

\author{N. L{\"o}rch}
\affiliation{Institute for Theoretical Physics and Institute for Gravitational Physics (Albert-Einstein-Institute), Leibniz University Hannover, Callinstrasse 38, 30167 Hannover, Germany}
\affiliation{Department of Physics, University of Basel, Klingelbergstrasse 82, 4056 Basel, Switzerland}

\author{I. D. Leroux}
\affiliation{QUEST Institut, Physikalisch-Technische Bundesanstalt, 38116 Braunschweig, Germany}

\author{P. O. Schmidt}
\affiliation{QUEST Institut, Physikalisch-Technische Bundesanstalt, 38116 Braunschweig, Germany}
\affiliation{Institute for Quantum Optics, Leibniz University Hannover, Welfengarten 1, 30167 Hannover, Germany}

\author{K. Hammerer}
\affiliation{Institute for Theoretical Physics and Institute for Gravitational Physics (Albert-Einstein-Institute), Leibniz University Hannover, Callinstrasse 38, 30167 Hannover, Germany}

\begin{abstract}
Optical clocks based on ensembles of trapped ions promise record frequency accuracy with good short-term stability. Most suitable ion species lack closed transitions, so the clock signal must be read out indirectly by transferring the quantum state of the clock ions to cotrapped logic ions of a different species. Existing methods of quantum logic readout require a linear overhead in either time or the number of logic ions. Here we describe a quantum algorithmic readout whose overhead scales logarithmically with the number of clock ions in both of these respects. The scheme allows a quantum nondemolition readout of the number of excited clock ions using a single multispecies gate operation which can also be used in other areas of ion trap technology such as quantum information processing, quantum simulations, metrology, and precision spectroscopy.
\end{abstract}

\date\today

\maketitle

Tremendous progress has recently been made in optical frequency metrology \cite{poli_optical_2013, Ludlow2015}. Optical clocks now reach fractional frequency inaccuracies and instabilities in the $10^{-18}$ regime \cite{rosenband2008, chou_frequency_2010, chou_quantum_2011, hinkley_atomic_2013, bloom_optical_2014, ushijima_cryogenic_2015}, outperforming Cs fountain clocks by two orders of magnitude and vying to serve as a new definition of the SI second \cite{gill_when_2011, riehle_towards_2015}. Among the promising candidates for such a redefinition are ion-based frequency standards featuring very small systematic frequency shifts. However, the poor signal-to-noise ratio of single-ion systems entails averaging times of many weeks to reach a fractional uncertainty of $10^{-18}$ \cite{santarelli_frequency_1998, peik_laser_2006}. Clocks based on strings of ions confined in a linear trap promise to overcome this limitation \cite{herschbach_linear_2012, pyka_high-precision_2014}. Because of unavoidable electric field gradients in such a trap, suitable clock ion species may have only negligible electric quadrupole moments to avoid systematic frequency shifts \cite{itano_external-field_2000}.
This requirement is met by group 13 ion species \cite{dehmelt_coherent_1981}, 
as well as some highly charged ions \cite{schiller_hydrogenlike_2007, derevianko_highly_2012, dzuba_ion_2012, dzuba_optical_2015}. Most of these candidates
lack a suitable transition for laser cooling and state detection, so that quantum logic spectroscopy \cite{schmidt_spectroscopy_2005} is required for readout. In quantum logic spectroscopy, the internal state of the clock ion is transferred by a series of laser pulses onto a logic ion of a different species, whose internal state can be efficiently detected. However, existing  methods for quantum logic readout require a large overhead in either time or logic ions when an ensemble of clock ions is used.

\begin{figure}[t]
  \centering
  \includegraphics [width=\columnwidth]{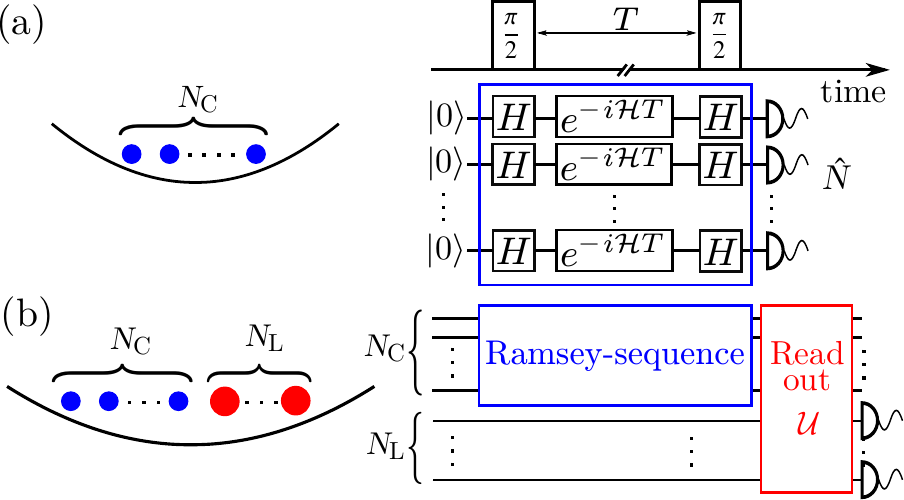}
  \caption{(a) In an atomic clock based on $\Nc$ trapped ions the frequency of a laser is locked to an atomic transition e.g., in a Ramsey sequence of two $\pi/2$ pulses (Hadamard gates $H$) enclosing a free evolution time $T$. Measurement of the number $\hat{N}$ of clock ions in $\ket{1}$ yields an error signal for the deviation of the laser from resonance. (b) For ion species lacking the cycling transition needed for direct state detection, a quantum algorithmic readout can be used to map $\hat{N}$ onto $\Nl$ cotrapped logic ions of a different species whose state can be detected efficiently.}\label{Fig:clock}
\end{figure}

Here, we suggest using a quantum algorithmic readout, cf. Fig.~\ref{Fig:clock}, which maps the clock signal (that is, the number of clock ions left in the excited state by the spectroscopy pulses) on a string of logic ions in \emph{binary} representation and therefore requires a number of logic ions scaling logarithmically with the number of clock ions.  Our approach is based on an algorithm originally introduced in the context of entanglement concentration protocols \cite{Bennett1996, Kaye2001}, which can be implemented efficiently with a number of elementary gate operations scaling likewise logarithmically with the number of clock ions. Moreover, we show that the algorithm can be executed by means of the multi-ion M{\o}lmer-S{\o}rensen (MS) gate \cite{Molmer1999,Solano1999,Milburn2000,Roos2008}, a well-established tool for quantum control of ion crystals \cite{Soderberg2010,Blatt2012}. In particular, we demonstrate that only a \emph{single} application of a MS gate involving both logic and clock ions is necessary. Such multispecies multi-ion gate operations are under intense experimental investigation at the moment \cite{Tan2015_nat}, and constitute the most costly resource in the current context due to the complexity of the normal mode structure in multispecies ion crystals. We discuss the feasibility of our approach in a case study of 3 \Al and 2 \Ca ions taking into account the full spectrum of motional normal modes.

The suggested readout method actually realizes a quantum nondemolition (QND) measurement on the clock ions, and therefore opens up the possibility to devise more sophisticated clock protocols where subsets of clock ions are interrogated repeatedly. Beyond frequency metrology the QND character of our readout technique makes it directly applicable to syndrome measurements and error correction schemes for quantum computations and simulations with single- or multispecies ion crystals \cite{Home2013}. In a more general perspective, this work contributes to the development of hybrid systems striving to combine the complementary strengths of different physical components for quantum information processing, quantum simulations and quantum metrology \cite{Wallquist2009, Kurizki2015}.

 \begin{figure*}[t]
  \centering
   \includegraphics [width=\textwidth]{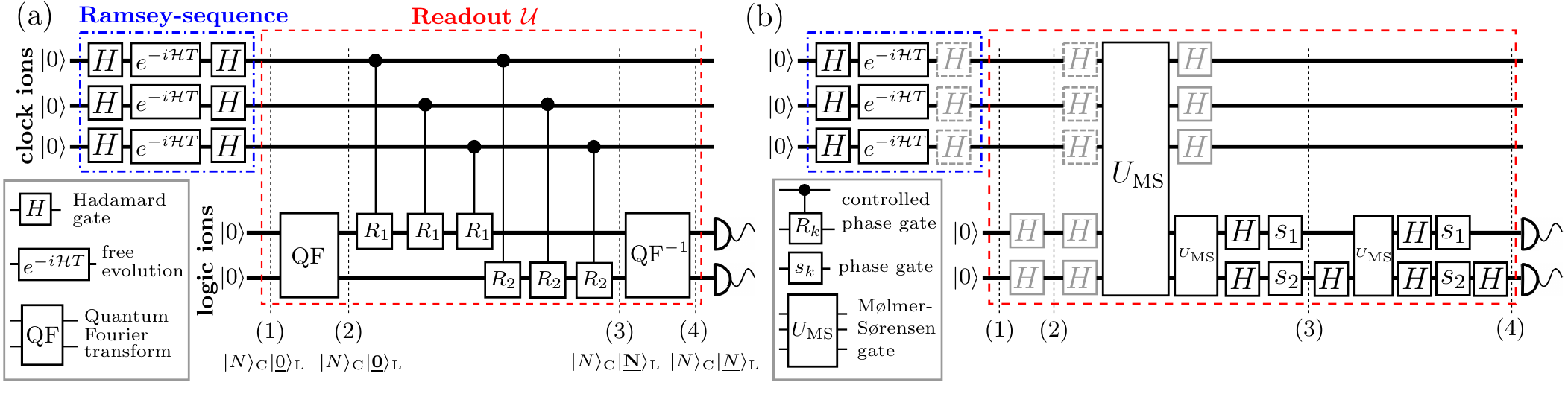}
    \caption{Quantum algorithmic readout illustrated for $\Nc=3$ clock and $\Nl=2$ logic ions: after a Ramsey sequency (blue, dash-dotted box) the clock ions are in a superposition of states $\ket{N}_\C$, where $N\in[0,\Nc]$ is the number of ions in excited states $\ket{1}$ (their Hamming weight). A quantum algorithmic readout $\mathcal{U}$ (red, dashed box)  maps $N$ in binary representation onto a number $\Nl=\lceil\log_2(\Nc+1)\rceil$ of logic ions. Detection of the logic ions in the $\{\ket{0},\ket{1}\}$ basis provides the number $N$. (a) Quantum algorithm  for the indirect measurement of the Hamming weight, taken from Refs. \cite{Bennett1996,Kaye2001}. (1) Logic ions are initialized in $\ket{\underline{0}}_\L=\ket{00}_\L$ (states $\ket{\underline{N}}_\L$ denote the binary representation of $N$). (2) The state is quantum Fourier transformed into $\ket{\underline{\mathbf{0}}}_\L$. (3) Controlled-phase gates rotate the state in Fourier space to $\ket{\underline{\mathbf{N}}}_\L$ for a state $\ket{N}_\C$ of clock ions. (4) An inverse Fourier transform yields the state $\ket{\underline{N}}_\L$ of logic ions. (b) The same algorithm decomposed in terms of multi-ion M{\o}lmer-S{\o}rensen (MS) gates. The Hadamard gates shown in gray need not be executed. Removing Hadamard gates displayed with dashed boxes merges  the algorithmic readout with the Ramsey sequence, and requires the laser fields in the MS gates to be phase coherent with the Hadamard gates in the Ramsey sequence. Explicit, general forms of all gates are given in the appendix.}\label{Fig:QAlg}
\end{figure*}

\paragraph*{Working principle of ion clocks with direct readout---} We consider a string of $\Nc$ clock ions with a narrow-band optical transition of frequency $\omega_0$ between two internal states $\ket{0}$ and $\ket{1}$ which provides the frequency reference for the clock. The goal is to stabilize to this transition frequency $\omega_0$ a laser field of frequency $\omega$ with an (unknown) frequency offset $\Delta=\omega-\omega_0$.  To this end, pulses of light from the laser drive the clock ions, transferring them from $\ket{0}$ to the excited state $\ket{1}$ with a frequency-dependent probability.  In the simplest schemes, such as Rabi and Ramsey interrogation, this probability is independent for each ion.  The clock readout then consists of measuring the number of excited ions $\hat{N}=\sum_{i=1}^{\Nc}\ket{1}_i\bra{1}$ and using it to infer the excitation probability and hence the detuning $\Delta$, which can then be corrected  \footnote{For definiteness we illustrate our proposed scheme with Ramsey interrogation (cf. Fig.~\ref{Fig:QAlg}a), but note that arbitrary clock protocols involving entangled states and correlations between ion excitation probabilities can always be designed to yield $\Nc+1$ measurement eigenvalues corresponding to the different possible numbers of excited ions~\cite{Buzek1999}.  A method to measure $\hat{N}$ can thus be used to read out any $\Nc$-ion clock.}; cf. Fig.~\ref{Fig:clock}(a). As explained above, a direct measurement of the excited state population $\hat{N}$ is impractical for many interesting species of clock ion. Instead, one can map $\hat{N}$ onto an ensemble of $\Nl$ cotrapped logic ions which can be detected efficiently, as illustrated in Fig.~\ref{Fig:clock}(b). In direct extension of established readout techniques based on quantum logic \cite{schmidt_spectroscopy_2005,Rosenband2007}, one could use as many logic ions as clock ions ($\Nl=\Nc$), performing successive quantum gate operations to transfer the state of each clock ion to the corresponding logic ion. Alternatively, one could use a single logic ion, transferring and measuring the state of each clock ion in turn. This imposes a long readout time as the ion crystal must be cooled between each measurement and subsequent state swap operation. Both strategies have prohibitive overhead in additional ions or number of gate operations and readout time.  The last is crucial because time spent on readout adds to the clock cycle's dead time and, thus, through the Dick effect, to the clock instability~\cite{Dick1987,Dick1990}.

\paragraph*{Quantum algorithmic readout---}
From the perspective of quantum information theory the quantity of interest -- the number of clock ions in state $\ket{1}$ -- is the \emph{Hamming weight} of the string of $\Nc$ quantum bits (a number with $\Nc+1$ possible values between 0 and $\Nc$). In the context of entanglement concentration protocols, a quantum algorithm has been developed for the indirect determination of the Hamming weight of a quantum bit string \cite{Bennett1996, Kaye2001}. The algorithm uses an ancillary string of $\Nl=\lceil\log_2(\Nc+1)\rceil$ quantum bits on which the Hamming weight of the $\Nc$ primary quantum bits is stored in binary representation. Thus, the necessary number $\Nl$ of logic ions (ancillary quantum bits) scales logarithmically with the number of clock ions. Suitable combinations of clock and logic ion numbers $(\Nc,\Nl)$ are, for example, $(3,2)$, $(7,3)$, and $(15,4)$. The algorithmic readout is shown in a circuit diagram for the simple case $(\Nc,\Nl)=(3,2)$ in Fig.~\ref{Fig:QAlg}(a), and for the general case in the appendix.  Given $\Nc$ quantum bits (clock ions) in any state $\ket{N}_\C $ with Hamming weight $N$ and $\Nl$ ancillary bits (logic ions) initialized in $\ket{00\ldots 0}_\L$, the algorithm effects a unitary transformation $\mathcal{U}$ such that $\mathcal{U}\ket{N}_\C \ket{00\ldots 0}_\L=\ket{N}_\C \ket{i_1i_2\ldots i_{\Nl}}_\L$, where the bit string $i_1i_2\ldots i_{\Nl}$ ($i_n=0,1$) gives the binary representation of the Hamming weight of the state of clock ions, $N=\sum_{n=1}^{\Nl}2^{n-1} i_n$. We will denote by  $\ket{\underline{N}}_L=\ket{i_1i_2\ldots i_{\Nl}}_\L$ the state of logic ions representing $N$. The state of the clock ions after one clock cycle is a superposition $\ket{\psi(\Delta)}_\C=\sum_{N=0}^{\Nc}c_N(\Delta)\ket{N}_\C$, where the dependence of the amplitudes $c_N(\Delta)$ on the detuning carries the clock signal \footnote{Without loss of generality we can assume that {$\ket{N}_\C$} denotes the normalized, symmetric superpostion of {$N$} clock ions in {$\ket{1}$} and all others in {$\ket{0}$}, that is, {$\ket{N}_\C=\binom{N}{\Nc}^{-1/2}J_+^N\otimes_{i=1}^{\Nc}\ket{0}$} where {$J_+=\sum_{i=1}^{\Nc}\sigma_+^i$}}.
Application of the readout algorithm generates an entangled state of clock and logic ions:
\[
\ket{\Psi(\Delta)}=\mathcal{U}\ket{\psi(\Delta)}_\C\ket{\underline{0}}_\L=\sum_{N=0}^{\Nc}c_N(\Delta)\ket{N}_\C\ket{\underline{N}}_L.
\]
From a measurement in the $\{\ket{0},\ket{1}\}$ basis of each logic ion one can extract the observable corresponding to the \emph{estimated} Hamming weight of the clock ions, $\hat{N}_\mathrm{est}=\sum_{n=1}^{\Nl}2^{n-1} \ket{1}_n\bra{1}=\sum_{N=0}^{\Nc} N\ket{\underline{N}}_{\L}\bra{\underline{N}}$. In a perfect implementation the measurement of $\hat{N}_\mathrm{est}$ on the logic ions exhibits \emph{exactly} the same statistics as measuring $\hat{N}$ on the clock ions directly. In particular, the error signal can be extracted from $N(\Delta)=\braket{\Psi(\Delta)|\hat{N}_\mathrm{est}|\Psi(\Delta)}$ and used to correct $\omega$ exactly as in a direct readout.

\begin{figure}[tb]
  \centering
   \includegraphics [width=\columnwidth]{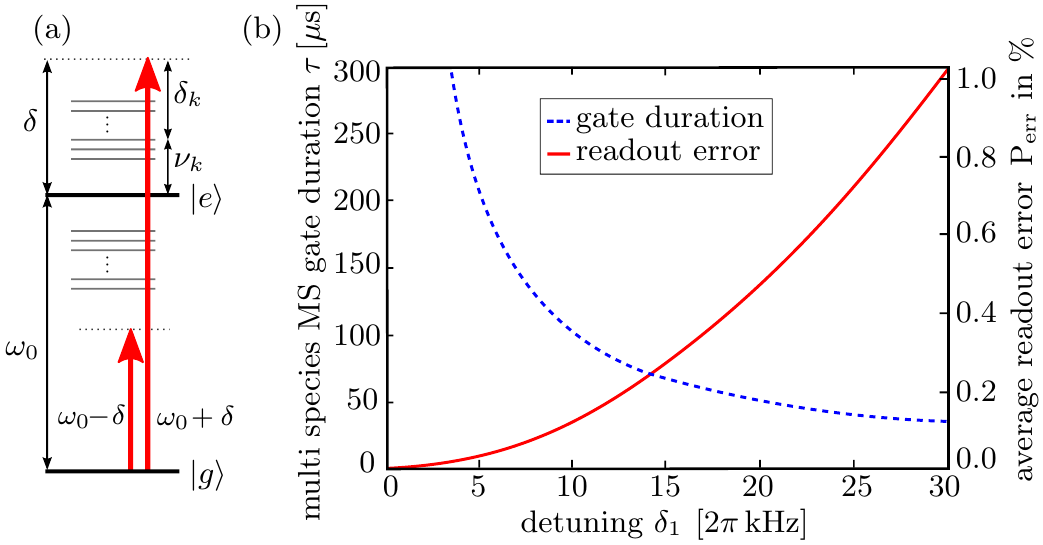}
    \caption{(a) Schematic level diagram with relevant frequencies during MS gates: atomic transition $\omega_0$, bichromatic driving fields at $\omega_0\pm\delta$, detuning $\delta_k$ from sideband $\nu_k$ due to collective mode $k$. (b) Average readout error probability $ P_{\mathrm{err}} $ (red, solid line) and multispecies MS gate duration $\tau$ (blue, dashed line) versus detuning $\delta_1$.
    }\label{Fig:ErrorProb}
\end{figure}

\paragraph*{Implementation based on M{\o}lmer-S{\o}rensen gates---} Following Refs. \cite{Bennett1996, Kaye2001}, the operation $\mathcal{U}$ can be implemented by a sequence of quantum Fourier transforms and controlled-phase gates as shown in Fig.~\ref{Fig:QAlg}(a). This requires a number of gates linear in $\Nc$, and so provides little advantage over the complete state swap mentioned above (apart from the reduction in the number of logic ions). Fortunately, the algorithm can be decomposed much more efficiently using multi-ion M{\o}lmer-S{\o}rensen (MS) gates \cite{Molmer1999,Solano1999,Milburn2000}. MS gates involve driving the ions  simultaneously with bichromatic laser fields on red and blue sideband transitions to collective modes of vibrations for a duration $\tau$, and generate unitary transformations $U_\mathrm{MS}=\exp[-i(S_{\C\C}+S_{\L\L}+S_{\C\L})]$ on the string of $N_\C+N_\L$ two-level systems where \cite{Zhu2006}
\begin{align}\label{eq:S}
S_{\alpha\beta}&=\sum_{i=1}^{N_\alpha} \sum_{j=1}^{N_\beta} \frac{\Gamma^{\alpha\beta}_{ij}}{\Delta^{\alpha\beta}_{ij}}\sigma^x_{\alpha i}\sigma^x_{\beta j}.
\end{align}
Here, $\sigma^x_{\alpha i}$ denotes the Pauli $x$ operator for ion $(\alpha, i)$, where $\alpha=\C, L$ and $i=1,\ldots,N_\alpha$, and the frequencies $\Gamma^{\alpha\beta}_{ij}$ and $\Delta^{\alpha\beta}_{ij}$ are defined through
\begin{align}\label{Eq:Delta}
  \Gamma^{\alpha\beta}_{ij}&=\int_0^\tau\!dt\, \Omega_{\alpha i}(t)\Omega_{\beta j}(t), & \frac{1}{\Delta^{\alpha\beta}_{ij}}&=\sum_{k=1}^{N_\L+N_\C}\frac{\eta_{\alpha i}^k\eta_{\beta j}^k}{\delta_k}.
\end{align}
We assume that the Rabi frequencies $\Omega_{\alpha i}(t)$ can be adjusted separately for each ion in transverse illumination, and may also be tuned in time. In the definition of $\Delta^{\alpha\beta}_{ij}$ in Eq.~\eqref{Eq:Delta} the $\eta^k_{\alpha i}$ are Lamb-Dicke factors and $ \delta_k $ detunings with respect to the $k$ th transverse normal mode of (angular) frequency $\nu_k$; cf. Fig.~\ref{Fig:ErrorProb}(a).
The detunings have to satisfy $\delta_k \gg \vert \Omega_{\alpha i}\eta^k_{\alpha i} \vert$ and we suppose $ \delta_1 > 0 $.
A decomposition of the desired transformation $\mathcal{U}$ in terms of multi-ion MS gates and single-ion rotations is shown in Fig.~\ref{Fig:QAlg}(b). Remarkably, only a \emph{single} two-species MS gate is required when the interaction between clock and logic ions, described by $S_{\C\L}$ in Eq.~\eqref{eq:S}, is tuned such that
\begin{equation}
  \Gamma^{\C\L}_{ij}=\pi\, 2^{-(j+2)}\Delta^{\C\L}_{ij} \label{eq:cond1}
\end{equation}
where $i=1,\ldots,\Nc$, $j=1,\ldots,\Nl$ \footnote{We assume $\Delta^{\alpha\beta}_{ij}\neq 0$, which is true for generic values of the detuning $\delta$.}. Eq.~\eqref{eq:cond1} ensures that the MS gate executes the controlled phase gates between logic and clock ions as shown in Fig.~\ref{Fig:QAlg}(a).
Read as a matrix equation, it expresses $\Nc\times\Nl$ constraints where the right-hand side is fixed for a given configuration of the ion string and carrier detuning $\delta = \delta_k + \nu_k $, while the left-hand side has to be adapted accordingly through the correct choice of Rabi frequencies $\Omega_{\alpha i}(t)$ and the gate duration $\tau$. In addition to the interactions between clock and logic ions ($S_{\C\L}$), the gate also induces interactions among clock ions ($S_{\C\C}$) and logic ions ($S_{\L\L}$) only. The interactions on the clock ions have no influence on the readout results and can therefore be neglected. On the other hand, interactions between logic ions have to be compensated using $\Nl-1$ gates that involve only logic ions, as shown in the appendix. The inverse Fourier transform on the string of logic ions involves $\Nl-1$ single-species MS gates giving a total number $ 2 \Nl - 1$ of multiqubit gates (see Appendix) that grows \emph{logarithmically} with the number of clock ions $\Nc$.

The remaining challenge is to find experimental parameters that satisfy Eq.\,\eqref{eq:cond1}. If each ion can be addressed individually, the time dependence of $\Omega_{\alpha i}(t)$ can be exploited to exactly fulfill Eq.~\eqref{eq:cond1}; as shown in the appendix. Here, we will instead focus on finding an approximate solution to condition~\eqref{eq:cond1} using only time-independent Rabi frequencies. This restriction substantially simplifies the experimental implementation, which can then employ a single laser beam for each species with a tailored intensity profile, rather than requiring full individual addressing with independent pulse shaping for each ion. In this case the left-hand side of Eq.~\eqref{eq:cond1} becomes $\Gamma^{\C\L}_{ij}= \Omega_{\C i}\Omega_{\L j}\tau$, which is patently a rank-one matrix while the rhs of Eq.~\eqref{eq:cond1} generically has full rank ($\Nl$). However, if the detuning $ \delta_1$  is small on the scale of the mode spacing $\Delta \nu = \nu_1- \nu_2 $, the ion interactions in the MS gate are mostly mediated via the highest-frequency collective mode. In this regime the rhs of Eq.~\eqref{eq:cond1} will be almost of rank one in the sense that its singular value decomposition exhibits one dominating singular value. Accordingly, the rhs of Eq.~\eqref{eq:cond1} can be well approximated in terms of a rank-one matrix by determining its singular value decomposition and setting all but the largest singular value to zero as detailed in the appendix. We emphasize that this is not equivalent to simply dropping lower-lying modes [which would correspond to restricting the summation in Eq.~\eqref{Eq:Delta} to $k=1$].

%

\paragraph*{Case study with \Al and \Ca ions---}
In the remainder of this Letter we discuss in more detail the realization of multispecies MS gates with constant Rabi frequencies for the concrete case of three \Al clock ions and two \Ca logic ions.
We assume the ions are held in a linear rf Paul trap with soft confinement along the crystal axis ($z$ axis) and much tighter confinements in the transverse directions that effectively restrict oscillations to only one direction ($x$ axis).
An important requirement for the MS gates is truly collective normal modes; i.e., each ion must be involved in the motion to interact with the others. For the case of multispecies crystals, these modes are found in a stable regime close to the critical confinement where the linear chain changes to a zigzag configuration \cite{Home2013,Kielpinski2000,Morigi2001}. In the appendix we show for the  case of \Al-\Ca crystals that for moderate trap asymmetry the highest-frequency mode is sufficiently collective, and exhibits Lamb-Dicke factors on the order of $10^{-1}$ for clock ions and $10^{-3}  - 10^{-2}$ for logic ions while exhibiting a substantial frequency gap $\Delta\nu\simeq 2\pi\,500$ kHz to the next mode. The approximate solutions to Eq.~\eqref{eq:cond1} will be applicable when the detuning in the multispecies MS gate is small enough, that is, $\delta_1 \ll \Delta\nu$. For a given detuning $\delta_1$, we require a gate duration $\tau=2\pi /\delta_1$ in order to suppress spurious effects of off-resonant terms during the MS gate, which implies a time scale in the sub-ms regime for the gate.

 In order to assess the performance of our readout strategy, we evaluate the average error probability defined as $P_{\mathrm{err}}= 1 - \sum_{N=0}^{\Nc} p_N \, P \left[N \vert \rho_{C}(N) \right] $. Here  $P \left[N \vert \rho_{C}(N) \right]$ is the conditional probability to correctly receive the Hamming weight $N$ as a result of the quantum algorithmic readout given that exactly $N$ ions were in the excited state in the initial state of clock ions $ \rho_{C}(N)$. The $p_N$ denote the \textit{a priori} probabilities for the states $ \rho_{C}(N) $. In the context of ion clocks, the maximum error signal will be attained when each clock ion is in an equal superposition of $|0\rangle$ and $|1\rangle$, which implies that $ p_N= \binom{\Nc}{N}/2^{\Nc} $ and $ \rho_{C}(N)$ is the symmetric Dicke state with $N$ clock ions in $\ket{1}$. The conditional probabilities are evaluated as $P \left[N \vert \rho_{C}(N) \right]=\mathrm{tr}_{CL}\left\{|\underline{N}\rangle_L\langle\underline{N}|\,\mathcal{U}\rho_{C}(N)\otimes |00\rangle_L\langle 00|\mathcal{U}^\dagger\right\}$, where $\mathcal{U}$ is the circuit shown in Fig.~\ref{Fig:QAlg}(b).  For a given detuning $\delta_1$ the Rabi frequencies in the single multispecies MS gate follow from the vectors corresponding to the largest singular value of the rhs of Eq.~\eqref{eq:cond1}, as explained above. Figure~\ref{Fig:ErrorProb} shows the overall error probability $P_{\mathrm{err}}$ and the gate duration $\tau$ versus detuning $\delta_1$ in the MS gate with otherwise ideal processes. For example, a detuning $\delta_1=2\pi\,20$\,kHz corresponds to a gate duration of $\tau= 50\, \mu \mathrm{s} $ and an error probability of $0.5\%$, which is achieved with Rabi frequencies $(\Omega_{\L1},\,\Omega_{\L2})=2\pi\,(730.34,\,363.81)$\,kHz and  $(\Omega_{\C1},\,\Omega_{\C2},\,\Omega_{\C3})=2\pi\,(49.71,\,45.84,\,53.81)$\,kHz for logic and clock ions, respectively. In this example the carrier transition is driven off resonantly with a Rabi frequency to detuning ratio of at most $\Omega_{\alpha i}/\delta = 0.23$. However, the associated ac Stark shifts cancel due to the bichromatic, red- and blue-detuned drive and can therefore be neglected. At the same time, the $50$~$\mu$s gate duration is short enough compared to the $1.17$~s $D_{5/2}$ state lifetime in \Ca that readout errors due to spontaneous emission will be on the per mill level. The Rabi frequencies required in the present example can be tailored by implementing the gate with a tightly focused $\mathrm{TEM}_{10}$ laser beam addressing the three clock ions, such that each ion is located at a position in the transverse intensity profile corresponding to the Rabi frequency given above.

 The more complex spatial structure of Rabi frequencies needed in a longer string of ions can be engineered with spatial light modulators \cite{mcgloin_applications_2003} or multichannel acousto-optical modulators. Combined with the freedom to permute clock and logic ions and to choose different solutions to Eq.~\eqref{eq:cond1}, we expect experimentally feasible implementations for more than 15 clock and 4 logic ions. Importantly, readout errors will affect only the stability, not the accuracy, of the ion clock.
\paragraph*{Conclusion---} The quantum algorithmic readout suggested here performs a QND measurement of the Hamming weight of clock ions and may therefore allow more complex clock protocols using repeated readouts of (sub)ensembles of ions or preparation of the clock ions in Dicke states for nonclassical frequency metrology. We envision that the techniques developed here in the context of atomic clocks may prove useful also in other areas of ion trap technology, such as quantum information processing, quantum simulations, metrology, and precision spectroscopy.

We acknowledge support from DFG through QUEST. I. D. L. acknowledges support from the Alexander von Humboldt Foundation. This work was supported by the European Metrology Research Programme (EMRP) in project SIB04. The EMRP is jointly funded by the EMRP participating countries within EURAMET and the European Union.

\bibliography{IonClocks}

\onecolumngrid

\subsection{Appendix}

\subsection{A.~~~Calculation of normal mode spectrum for transversal oscillations in a multi-species ion crystals}

\begin{figure}[htb]
 \centering
  \includegraphics [width=0.7 \columnwidth]{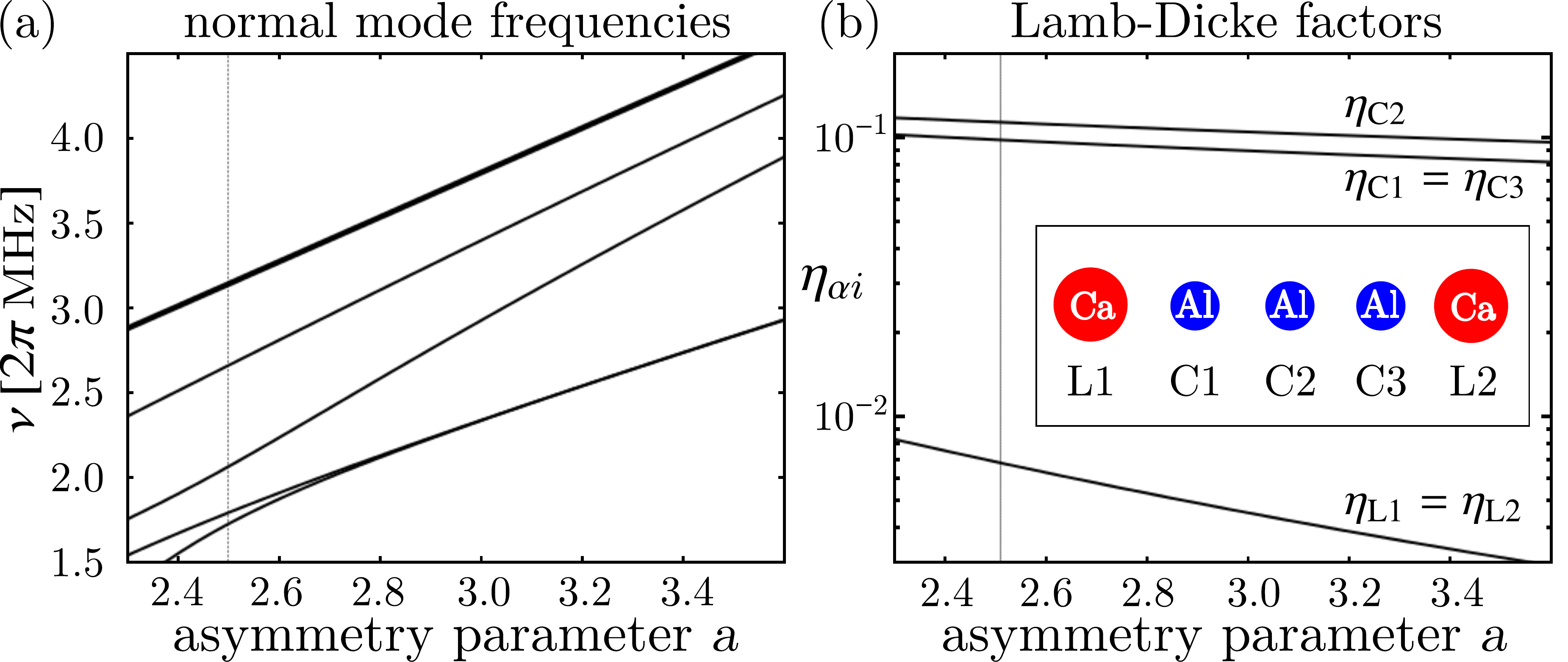}
   \caption{(a) Normal mode frequencies of transverse vibration for a crystal of $\Nc=3$ Al$^+$ clock ions between $\Nl=2$ Ca$^+$ logic ions (see b, inset) versus asymmetry parameter $a=\nu_x^\L/\nu_z^\L$ (ratio of transverse to axial trap frequencies for logic ions, $\nu_z^\L=2\pi\,874$\,kHz). (b) LD factors $\eta_{\C i}^1$ and $\eta_{\L j}^1$ of, respectively, the three clock and two logic ions for the normal mode of highest frequency (thick line in (a)). For large asymmetry parameter transverse motions of clock and logic ions decouple due to the mass-dependent transverse confinement. The dotted line in (a) and (b) marks the case studied in the text.}\label{Fig:NormalModes}
\end{figure}

For sake of completeness we summarize the determination of the normal mode spectrum for the transversal oscillations of a two-species ion chain, following the treartment of \cite{Morigi2001}. In the example setup described in the main text two logic ions (\Ca, mass $m_L=40 \, \mathrm{amu}$, laser wavelength $\lambda_L=729.1\, \mathrm{nm}$) and three clock ions (\Al, mass $m_C=27 \, \mathrm{amu}$, laser wavelength $\lambda_C=267.4 \, \mathrm{nm}$) are trapped along the crystal axis ($z$-axis). The trap frequency for the logic ions is   $\nu_{z}^L=2\pi 874 \, \mathrm{kHz}$. The asymmetry parameters are defined as $a=\nu_{x}^L/\nu_{z}^L$ and $a_{yx}=\nu_{y}^L/\nu_{x}^L$. We fix $a_{yx}=5$ to suppress oscillations along the $y$-axis. The normal modes are calculated for different values of  $a$, because the oscillations in $x$-directions will be used for the gate. As sketched in Fig.~\ref{Fig:NormalModes}(b) the clock ions (index $k=2, 3, 4$) are placed in the middle and the logic ions (index $k=1$ and $k=5$) on the outside. With masses $(m_1, m_2, m_3, m_4, m_5)=(m_L,m_C,m_C,m_C,m_L)$ the kinetic energy is given by
\begin{align*}
T(\vec p)= \sum_{k=1}^5\vec p_k^{\,2}/2m_k.
\end{align*}
The potential energy due to the electrostatic component of the trap is the same for each ion, as all ions carry a single positive elementary charge $q$. Assuming a radially symmetric electrostatic trap the total electrostatic potential is \cite{Kielpinski2000}
\begin{align*}
V_S(\vec x)=\tfrac 12   \sum_{k=1}^5 \left(b_0 z_k^2 -\tfrac 12 {b_0} x_k^2  -\tfrac 12 {b_0} y_k^2 \right),
\end{align*}
The potential strenth is parametrized by the parameter $b_0$, in units of energy divided by length squared. To keep the ions also radially trapped, an additional time-dependent radiofrequency potential is used to create an effective mass-dependent potential in $x$ and $y$ direction
\begin{align*}
V_{RF}(\vec x)=\tfrac 12   \sum_{k=1}^5 \frac{m_L}{m_k} \left( b_{x}  x_k^2 +b_{y} y_k^2 \right).
\end{align*}
The potential strength is parametrized by $b_x$ and $b_y$, in the same units as $b_0$. The parameters $b_0$, $b_x$ and $b_y$ are fully determined by the physical parameters of the setup: logic ion mass $m_L$, logic ion frequency $\nu_z^L$ and the asymmetry parameters $a$ and $a_{xy}$.
It is important to note that the two species of ions experience different radial potentials because the pseudopotential generated by the radial AC fields in a Paul trap is mass-dependent.
As a result, lighter ions feel a tighter transverse potential.
The corresponding trap frequencies of clock and logic ions are denoted by $\nu_x^\C$ and $\nu_x^\L$, where $\nu_x^\L<\nu_x^\C$ since $m_\mathrm{Al}/m_\mathrm{Ca}=27/40$. In addition to the trapping potential the ions interact via the Coulomb repulsion potential
\begin{align*}
V_I(\vec x)=\sum_{k>j} \frac{q^2}{4 \pi \epsilon_0} \left|\vec x_k - \vec x_j \right| ^{-1}.
\end{align*}
With these definitions we can write the total energy of the system as
$
E(\vec x, \vec p)=T(\vec p)+V(\vec x),
$
where $V(\vec x)=V_S(\vec x)+V_{RF}(\vec x)+V_I(\vec x)$.

We find the steady state position $\vec{x}_0$ of the ions by numerically minimizing \footnote{We use Powell's method as implemented in SciPy for the numerical minimization} the potential energy $V$ under the condition $z_1<z_2<...<z_5$. In this study we choose large enough asymmetry parameters $a$ so that the solution is always a linear chain without zigzag configuration \cite{Kielpinski2000,Morigi2001,Home2013}. As the oscillations around the steady state will be small, we use second order Taylor expansion to obtain an approximate harmonic potential. The different directions $x$, $y$ and $z$ decouple in this approximation. Denoting $p_k=m_k \dot x_k$ and $V_{kj}=\partial_{x_k x_j} V(\vec x) \bigg|_{\vec{x}_0}$ the energy for motion in $x$-direction only is
\begin{align*}
E_x=\sum_{k=1}^5 \frac{p_k^2}{2m_k} + \sum_{k,j} \frac{1}{2}  V_{kj} x_k x_j.
\end{align*}
In coordinates with scaled position $\tilde x_k=\sqrt{ \frac{m_k}{m_0}} x_k$ and momentum $\tilde p_k=\sqrt {\frac{m_0}{m_k}} p_k$, normalized to mass $m_0=1$~amu, the kinetic term becomes diagonal and the potential transforms as $\tilde V_{kj}=\sqrt {\frac{m_0^2}{m_k m_j}} V_{kj}$. In these coordinates $E_x$ reads
\begin{align*}
E_x=\sum_{k=1}^5 \frac{\tilde p_k^{\,2}}{2m_0} + \sum_{k,j} \frac{1}{2} \tilde V_{kj} \tilde x_k \tilde x_j.
\end{align*}
For the normal modes we numerically diagonalize $\tilde V = O D O^T$ with a dimensionless orthogonal matrix $O$ and diagonal matrix $D$ of dimension frequency squared times mass. The eigenfrequencies are then given by
\begin{align*}
 \nu_k=\sqrt{D_{kk}/{m_0}}.
\end{align*}
In analogy to \cite{Morigi2001} the Lamb-Dicke factors of a mode $k$ for an individual ion with index $j$ are
\begin{align*}
\eta_j^k=\frac{2 \pi}{\lambda_j} O_{j}^k \sqrt{\frac{\hbar}{2 m_j \nu_k}},
\end{align*}
with $\lambda_j$ the wavelength of the laser addressing the $j$-th ion and $O_{jk}$ the $j$-th entry of the eigenvector for the $k$-th normal mode.

In Fig.~\ref{Fig:NormalModes}(a) we show the spectrum of the transverse normal modes for the case of $(\Nc,\Nl)=(3,2)$ in the ordering shown in the inset of Fig.~\ref{Fig:NormalModes}(b).
The generic result is that for large asymmetry parameter $a$ the $\Nl+\Nc$ normal modes split into two groups involving either the $\Nl$ logic or the $\Nc$ clock ions.
To have truly collective normal modes involving both species of ions (as required for the M\o{}lmer-S\o{}rensen gate) the asymmetry must be kept moderate.
However, below a critical value of $a$ the normal modes become unstable and the crystal changes to a zig-zag configuration \cite{Home2013,Kielpinski2000,Morigi2001}.
For moderate trap asymmetry the highest lying modes is sufficiently collective, and exhibits Lamb-Dicke factors on the order of $10^{-1}$ for clock ions and $10^{-3}$ to $10^{-2}$ for logic ions while exhibiting a substantial frequency gap $\Delta\nu$ to the next mode.
For an axial trap frequency of $2\pi\,874$\,kHz and an asymmetry parameter $a=2.5$ the highest-lying transverse mode has a frequency $\nu_1=2\pi\,3.14$\,MHz and the gap to the next mode is $\Delta\nu=\nu_1 - \nu_2=2\pi\,480$\,kHz.
Assuming all laser beams are aligned with the $x$-axis, the corresponding LD factors for logic and clock ions are given in Tab.\ref{Tab:LD}.
The LD factors for the highest frequency mode are most relevant for the gate dynamics.
Slightly larger LD factors were used in \cite{Leibfried2003,Sackett2000} to achieve high-fidelity M\o{}lmer-S\o{}rensen (MS) gates on two ions and entanglement of up to four ions, respectively.
Other demonstrations of high-fidelity multi-ion gates used LD parameters which were a factor of two to three smaller than the values considered here \cite{Kim2009,Kirchmair2009,Monz2011}.
The information on the normal modes and the gate detuning is enough to calculate the r.h.s of Eq.~\eqref{eq:cond1} and find appropriate Rabi frequencies.

\begin{table}
\begin{tabular}{l|ccccc}
~	& $\L 1$ 	& $\C 1$	& $\C 2$	& $\C 3$	& $\L 2$ \\
\hline
$\nu_1=2\pi \, 3.14 \> \mathrm{MHz} $      & 0.007  & 0.098		& 0.113 & 0.098 & 0.007      \\
$\nu_2=2\pi \,  2.66 \>  \mathrm{MHz}$      & $-1.301 \cdot 10^{-2}$  & $-0.133$		& $-2.429 \cdot 10^{-8} $ & $0.133$ & $1.301 \cdot 10^{-2}$      \\
$\nu_3=2\pi \,  2.06 \>  \mathrm{MHz}$      & $-0.024$  & $-0.060$		& 0.137 & $-0.060$ & $-0.024$      \\
$\nu_4=2\pi \,  1.79 \>  \mathrm{MHz}$      & $3.996 \cdot 10^{-2}$  & $-4.318 \cdot 10^{-2}$		& $-1.079 \cdot 10^{-6}$ & $4.318 \cdot 10^{-2}$ & $-3.996 \cdot 10^{-2}$      \\
$\nu_5=2\pi \,  1.72 \>  \mathrm{MHz}$     & 0.034  & $-0.067$		& 0.071 & $-0.067$ & 0.034      \\
\hline
\end{tabular}
\caption{Numerical solutions of normal mode frequencies $ \nu_k $ and LD factors $ \eta_{\alpha i}^k $ for all ions and modes at $ a=2.5 $. Each row corresponds to a normal mode as indicated by the index of the corresponding frequency. The symmetry in LD factors is due to the symmetry of the ion configuration. }\label{Tab:LD}
\end{table}

\subsection{B.~~~Derivation of the effective M\o{}lmer-S\o{}rensen-Hamiltonian}

A M\o{}lmer-S\o{}rensen (MS) gate is achieved by the interaction of multiple ions with two laser fields equally detuned to the upper and lower sideband of a collective motional mode \cite{Molmer1999,Roos2008}. The two lasers $A$ and $B$ (red and blue detuned, respectively) have frequencies $ \omega_A = \omega_0 - \delta $ and $ \omega_B = \omega_0 + \delta $ where $ \delta $ is the detuning to the carrier frequency $ \omega_0 $.
Therefore $ \delta_k = \delta - \nu_k $ gives the detuning with respect to the motional mode $ \nu_k $ mediating the interactions.
Here $\delta_1 > 0 $ is supposed to be a small detuning to the motional sideband with highest frequency in Fig. \ref{Fig:NormalModes}(a).
First we assume a general setup of $N$ arbitrary ions with individual Lamb-Dicke factors $\eta_{A/B,j}^k$ for ion $j$ and collective mode $k$.
The bichromatic laser has, in general, time dependent Rabi frequencies $ \Omega_{A/B,j} (t) $ and a laser phase $\phi_{i}$ that is assumed to be equal for both lasers $A$ and $B$ but possibly different for each ion. The Hamiltonian of the system consists of the internal energies, motional energy and the interaction with the lasers. Changing into an interaction picture gives a time dependent Hamiltonian in Lamb-Dicke expansion ($\hbar=1$)
\begin{equation} \mathcal{H} (t) = \sum_{ j,k = 1}^{N} \Omega_{A ,j} (t) ~  \eta_{A, j}^k \left( e^{i \delta_k  t} e^{-i \phi_{j}} \sigma_{j}^+ a_k + \mathrm{h.c.} \right) + \Omega_{B , j}(t)~ \eta_{B, j}^k \left(e^{-i \delta_k t} e^{-i \phi_{j}} \sigma_{j}^+ a^{\dagger}_k  + \mathrm{h.c.} \right). \label{Eq:Ham}
\end{equation}
Here $a^{\dagger}_k$ and $a_k$ are the creation and annihilation operators of the collective motional mode with frequency $\nu_k$.
The Lamb-Dicke approximation was used keeping terms only to linear order in $\eta_{A/B,j}^k $ and a rotating wave approximation was applied, neglecting all terms rotating faster than $\delta_N $.  The laser phases can be absorbed through a unitary transformation $ \mathcal{H} (t) = V  \tilde{\mathcal{H}}  V^{\dagger}$ where
\[ V = \prod_{j = 1}^{N} \mathrm{exp} \left( -i \dfrac{\phi_{j}}{2}  \sigma_{j}^z \right) \]
such that
\[ \tilde{\mathcal{H}} (t) = \sum_{ j,k = 1}^{N} \Omega_{A , j} (t) ~  \eta_{A, j}^k \left(  e^{i \delta_k  t} \sigma_{j}^+  a_k + \mathrm{h.c.} \right) + \Omega_{B , j} (t) ~  \eta_{B, j}^k \left(  e^{-i \delta_k t} \sigma_{j}^+ a^{\dagger}_k + \mathrm{h.c.} \right). \]
The effective Hamiltonian can be calculated from $ \tilde{\mathcal{H}}(t) $ using a Magnus expansion
\[
U (\Delta t) 	=\mathrm{exp} \left( Y(\Delta t) \right) \]
with
\[ Y(\Delta t)	= -i\int_0^{\Delta t} \tilde{\mathcal{H}}(t) ~ dt + \dfrac{(-i)^2}{2} \int_0^{\Delta t} \left[ \int_0^{t} \tilde{\mathcal{H}}(t'') dt'' , \tilde{\mathcal{H}}(t) \right]  dt  + \ldots
\]
Oscillating integrands proportial to $ e^{i (\delta + \nu_1 - \nu_k ) t} $ or even faster phases are assumed to average to zero over the gate duration and therefore will be neglected. The unitary time evolution is then given by a time dependent effective Hamiltonian
\[ U (\Delta t) = \mathrm{exp} \left(-i ~ \tilde{\mathcal{H}}_{\mathrm{eff}} \left(  \Delta t \right) \right).
\]
and $\tilde{\mathcal{H}}_{\mathrm{eff}}$ can be written in a compact form as
\[\tilde{\mathcal{H}}_{\mathrm{eff}} \left( \Delta t \right) \approx \sum_{k=1}^N  \dfrac{1}{4 \delta_k } \int_0^{\Delta t} \left[ \tilde{S}_{x,k}^2 \left( t \right) + \tilde{S}_{y,k}^2 \left( t \right)  + \tilde{S}_{z,k} \left( t \right)  \right] dt
\]
where
\begin{align*}
\tilde{S}_{x,k} \left( t \right) &= \sum_{j=1}^N \left( \Omega_{A,j} \left( t \right)  \eta_{A,j}^k + \Omega_{B,j} \left( t \right)  \eta_{B,j}^k \right) \sigma_j^x,\\
\tilde{S}_{y,k} \left( t \right) &=  \sum_{j=1}^N \left( \Omega_{A,j} \left( t \right)  \eta_{A,j}^k - \Omega_{B,j} \left( t \right)  \eta_{B,j}^k \right) \sigma_j^y \\
\tilde{S}_{z,k} \left( t \right) &= \sum_{j=1}^N  2 \left( \Omega_{A,j} \left( t \right)  \eta_{A,j}^k + \Omega_{B,j} \left( t \right)  \eta_{B,j}^k \right)   \left( \Omega_{A,j} \left( t \right) \eta_{A,j}^k - \Omega_{B,j} \left( t \right)  \eta_{B,j}^k \right) \left( 2 a_k^{\dagger} a_k + 1 \right) \sigma_j^{z}
\end{align*}
This representation emphasizes the different contributions to the effective Hamiltonian. $\tilde{S}_{x,k}^2 $ and $\tilde{S}_{y,k}^2 $ give rise to the usual collective spin flips in a MS gate and $S_{z,k} $ are energy shifts for the internal states of the ions.
Note that both $\tilde{S}_{y,k}$ and $\tilde{S}_{z,k}$ are proportional to the differences in Rabi frequencies and Lamb-Dicke factors for the lasers $A$ or $B$ and therefore vanish if those are equal.
This Hamiltonian is an approximat solution assuming slowly varying Rabi-Frequencies compared to the detuning on the timescale of the gate duration.
Now the unitary transformation, with $V$ given above, is applied to $\tilde{\mathcal{H}}_{\mathrm{eff}} $ to find the full Hamiltonian of the MS interaction, namely
\[ \mathcal{H}_{\mathrm{MS}} \left( \Delta t \right) = V \tilde{\mathcal{H}}_{\mathrm{eff}} \left( \Delta t \right) V^{\dagger} = \sum_{k=1}^N  \dfrac{1}{4 \delta_k } \int_0^{\Delta t} \left[ \left( \tilde{S}_{x,k}^{\phi} \right)^2 \left( t \right) + \left( \tilde{S}_{y,k}^{\phi} \right)^2 \left( t \right) + \tilde{S}_{z,k} \left( t \right) \right] dt
\]
with the operators
\begin{align*}  \tilde{S}_{x,k}^{\phi} \left( t \right) = \sum_{j=1}^N \left( \Omega_{A,j} \left( t \right)  \eta_{A,j}^k + \Omega_{B,j} \left( t \right) \eta_{B,j}^k \right) \left( \sigma_j^x ~ \mathrm{cos} ~ \phi_j - \sigma_j^y  ~ \mathrm{sin}~ \phi_j 	\right)  \\
  \tilde{S}_{y,k}^{\phi} \left( t \right) = \sum_{j=1}^N \left( \Omega_{A,j} \left( t \right)  \eta_{A,j}^k - \Omega_{B,j} \left( t \right) \eta_{B,j}^k \right) \left( \sigma_j^y ~ \mathrm{cos}~ \phi_j + \sigma_j^x ~ \mathrm{sin}~ \phi_j 	\right)
\end{align*}
and $\tilde{S}_{z,k} \left( t \right)$ stays unchanged. 

If we assume now that each ion interacts with both laser beams in the same way, meaning that $\Omega_{A,j} \left( t \right)=\Omega_{B,j} \left( t \right) \equiv \Omega_j \left( t \right) $ and $\eta_{A,j}^k=\eta_{B,j}^k \equiv \eta_j^k $ hold for each ion $j$ and mode $k$, the effective Hamiltonian reduces to interactions given by the $ \tilde{S}_{x,k}^{\phi} \left( t \right) $ terms only:
\[ \mathcal{H}_{\mathrm{MS}} \left( \Delta t \right) = \sum_{k=1}^N  \dfrac{1}{ \delta_k} \int_0^{\Delta t} \left( \sum_{j=1}^N \Omega_j \left( t \right) \eta_j^k  \left( \sigma_j^x ~ \mathrm{cos} ~ \phi_j - \sigma_j^y  ~ \mathrm{sin} ~ \phi_j \right)\right) \left( \sum_{j'=1}^N \Omega_{j'} \left( t \right) \eta_{j'}^k  \left( \sigma_{j'}^x ~ \mathrm{cos} ~ \phi_{j'} - \sigma_{j'}^y  ~ \mathrm{sin}~ \phi_{j'}  \right)\right) dt
\]
Adjusting the Laser phases $\phi_j$ allows to alter between $ \sigma^x \sigma^x $ or $ \sigma^y \sigma^y $ interactions and also determines the overall sign of these. Choosing $ \phi_j \equiv \phi_{j'} = \pi $ simplifies the Hamiltonian to
\[ \mathcal{H}_{\mathrm{MS}} \left( \Delta t \right) = \sum_{j,j'=1}^N \sum_{k=1}^N  \dfrac{\eta_j^k \eta_{j'}^k}{ \delta_k } \sigma_j^x \sigma_{j'}^x \int_0^{\Delta t} \Omega_j \left( t \right) \Omega_{j'} \left( t \right) dt
\]

Finally, to compare this result to the quantum gate used in the main text, we define $ \Gamma_{jj'} \left( \Delta t \right) $ and $ \dfrac{1}{\Delta_{jj'}} $ as
\[  \Gamma_{jj'} \left( \Delta t \right) = \int_0^{\Delta t} \Omega_j \left( t \right) \Omega_{j'} \left( t \right) dt ~~~  \mathrm{and} ~~~  \dfrac{1}{\Delta_{jj'}} = \sum_{k=1}^N  \dfrac{\eta_j^k \eta_{j'}^k}{\delta_k }
\]
This gives
\[ U_{\mathrm{MS}} \left( \Delta t \right) = \mathrm{exp} \left( - i  S \left( \Delta t \right) \right) \]
 where
\[ S \left( \Delta t \right) = \sum_{j,j'=1}^N \dfrac{\Gamma_{jj'} \left( \Delta t \right)}{\Delta_{jj'}} \sigma_j^x \sigma_{j'}^x
\]
This is identical to the unitary evolution used in the readout strategy if we label the $N$ ions accordingly as clock or logic-ions, i.e $j \rightarrow \alpha i$.

\subsection{C.~~~Algorithmic Readout for General Number of Ions}\label{App:Readout}

\begin{figure}[htb]
\centering
\includegraphics[scale=0.7]{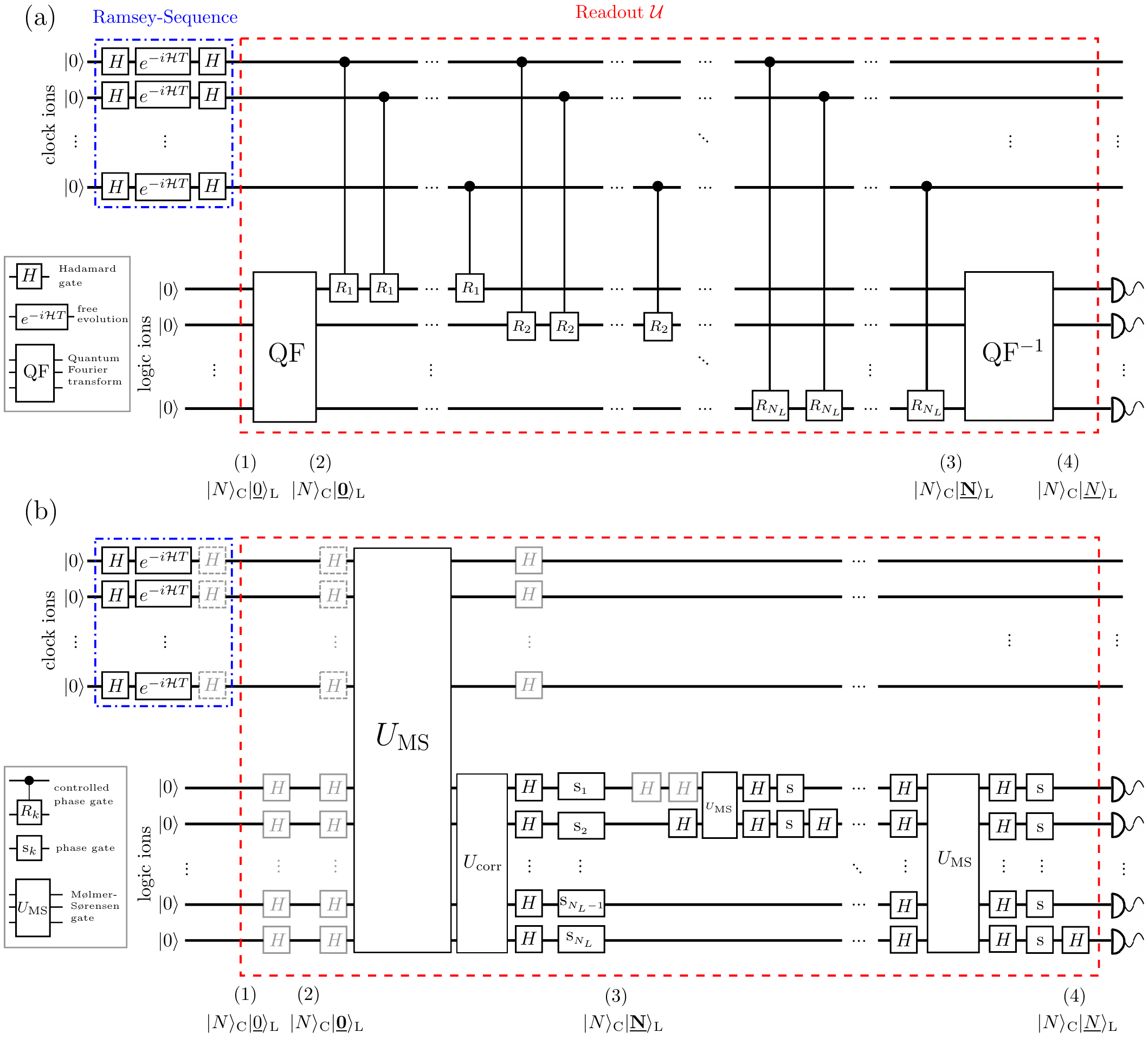}
\caption{Full algorithmic Readout for a general number of clock- and logic ions $N_{\mathrm{C}}$,$N_{\mathrm{L}}$.
(a) As described in the main text, the clock cycle starts with a Ramsey sequence (blue, dash-dotted box) on all $ N_{\mathrm{C}} $ clock ions, uses the quantum algorithm $ \mathcal{U} $ (red, dashed box) to encode the number of excited clock ions onto the $ N_{\mathrm{L}} $ logic ions and gives Hamming weight in binomial representation with a subsequent measurement of each logic ion.
So the unitary transformation  $ \mathcal{U} $ starts at (1) with the state $ \vert N \rangle_\mathrm{C} \vert \underline{0} \rangle_\mathrm{L} $ where $N$ clock ions are in the excited state and all logic ions are prepared in the ground state.
Then the quantum fourier transform is applied to the logic ions to give $  \vert N \rangle_\mathrm{C} \vert \underline{\textbf{0}} \rangle_\mathrm{L} $ as the resulting state at (2).
The controlled phase gates $ R_k $ 
add additional phases to the logic ions for each excited clock ion to give the state $  \vert N \rangle_\mathrm{C} \vert \underline{\textbf{N}} \rangle_\mathrm{L} $ at (3).
The inverse quantum fourier transform yields the state $  \vert N \rangle_\mathrm{C} \vert \underline{N} \rangle_\mathrm{L} $ at (4) so that the detection of each logic ion gives the binary representation of $N$.
(b) Decomposition into quantum gates for an experimental realisation.
In the first step the quantum fourier transformation is performed via Hadamard gates on the logic ions.
Then a M\o{}lmer-S\o{}rensen-gate connecting clock- and logic ions transfers the information about the Hamming weight onto the logic ions and the gates in $ U_{\mathrm{corr}} $ correct for unwanted dynamics.
The inverse quantum fourier transformation is decomposed into $ N_{\mathrm{L}} -1 $ M\o{}lmer-S\o{}rensen-gates and related single ion phase gates separated by a Hadamard gate.
For simplicity each phase gate is labeled $s$ although they all describe different phase shifts.
The detailed phases for $s$ and the coefficients for each $ U_{\mathrm{MS}} $ operation are given in the text.
Again, Hadamard gates shown in grey drop in pairs and do not need to be executed.
Grey dashed gates merge the readout with the Ramsey sequence.
 }\label{Fig:QAlgGeneral}
\end{figure}

Here we explain the readout algorithm -- which was discussed for the simple case  $ N_{\mathrm{C}} = 3 $ , $ N_{\mathrm{L}} = 2 $ in the main text -- for arbitrary number of clock and logic ions $\Nc$ and $\Nl$. The generalisation of the schematic circuit in Fig.~\ref{Fig:QAlg}(a) to the case of arbitrary numbers of ions (following \cite{Bennett1996, Kaye2001}) is shown in Fig.~\ref{Fig:QAlgGeneral}(a). The Ramsey-sequence is performed on all clock ions (leaving them in a superposition of states $\vert N \rangle_\mathrm{C}$), and the logic ions are prepared in the ground state at the beginning of the algorithm $ \mathcal{U} $. After applying the quantum fourier transformation on the logic ions the algorithm consists of controlled phase gates using the clock ions as control bits.
This way an excited clock ion gives an additional phase to the excited state in each logic ion via the unitary phase gate
\[
R_k = \begin{pmatrix}
1	&	0	\\
0	&	e^{2 \pi i/2^k}
\end{pmatrix}
\]
and generates the state $  \vert N \rangle_\mathrm{C} \vert \underline{\textbf{N}} \rangle_\mathrm{L} $. The inverse quantum fourier transformation produces the Hamming-weight $N$ in binary representation as the logic ions' state, which can then be retrieved by a measurement on the logic ions.

Generalising the implementation of the readout by means of M\o{}lmer-S\o{}rensen (MS) gates requires some more work than the schematic description, mostly due to the inverse quantum fourier transformation.
The initial quantum fourier transformation in $ \mathcal{U} $ is again given by Hadamard gates on the logic ions, since they were initially prepared in the ground state.
The main part of the algorithm, i.e. the controlled phase gates between clock- and logic ions, are performed
by a single MS gate and additional single qubit phase gates.
This MS gate is described in the main text and Eq.~\eqref{eq:cond1} shows the condition that the coefficients need to fulfill in order to implement the desired algorithm.
Additional to the desired dynamics a general setting will also involve MS interactions between all pairs of logic ions with interaction strengths $\Gamma^{\L\L}_{jj'}/\Delta^{\L\L}_{jj'}$, where $ j,j'=1, \cdots , \Nl $.
Applying a correction dynamic $ U_{\mathrm{corr}} $ can reverse these interactions.
Wherein $ U_{\mathrm{corr}} $ consists of $ \Nl - 1 $ M\o{}lmer-S\o{}rensen gates each involving a smaller subset of logic ions and the signs of the interactions are inverted by changing the laser phase for a different ion in each gate.
In Fig.~\ref{Fig:QAlgGeneral}(b) the single phase gates associated with $ U_{\mathrm{MS}} $ are labeled as $s_k$.
Those are controlled phase gates
\begin{align} \label{phasegate}
 s_k =  \begin{pmatrix}
1	&	0	\\
0	&	e^{i  \theta_k }
\end{pmatrix}
\end{align}
 with phases $ \theta_k =  N_{\mathrm{C}} ~ \pi ~ 2^{-k} $ for $ k = 1,...,N_{\mathrm{L}} $.
Additional single ion phase gates (and Hadamard gates) on the clock ions are left out in this discussion as well as in the given figures (displayed in grey).
These gates are only necessary to recover the initial state of the clock ions, but this is not the focus of the algorithmic readout presented here.
Hadamard transformations surrounding the MS gates are used to relate the physical $ \sigma_x \otimes \sigma_x$ to $\sigma_z \otimes \sigma_z$.

With more logic ions, the implementation of the inverse quantum fourier transformation become more costly in terms of the MS gates needed.
An efficient circuit for the quantum fourier transformation consists of blocks with different controlled phase gates separated by Hadamard gates, see e.g.~\cite{Nielsen2000}.
In analogy to other parts of the readout, these steps are performed using MS gates and single ion phase gates.
The inverse quantum fourier transformation on the logic ions is implemented by a series of $ N_{\mathrm{L}} -1 $ such steps, each involving an increasing number of ions and ending with a Hadamard gate on the last ion involved.
Every MS gate is described by the same mechanism as given by Eq.~\eqref{eq:S} in the main text.
It is therefore determined by $ \Gamma^{\L\L}_{jj'}/\Delta^{\L\L}_{jj'}$ on the logic ions.
For the step involving ions $1$ to $k$ the coefficients are

\[
\dfrac{\Gamma^{\L\L}_{jj'}}{\Delta^{\L\L}_{jj'}} = - \pi \cdot 2^{j+j'-3}  ~~~~ \mathrm{and}  ~~~~ \dfrac{\Gamma^{\L\L}_{jk}}{\Delta^{\L\L}_{jk}} = - \pi \cdot 2^{j-k-1}
\]
with $ j,j'=1,2,...,k-1 $.
For $ k = N_{\mathrm{L}}  $ the largest coefficient  $ \dfrac{\Gamma^{\L\L}_{N_{\mathrm{L}}-1, N_{\mathrm{L}}-2  }}{\Delta^{\L\L}_{N_{\mathrm{L}}-1, N_{\mathrm{L}}-2  }} = - \pi \cdot 2^{2 N_{\mathrm{L}} -6 } $ gives the extreme case, requiring the largest Rabi frequencies. This can set a limit to possible implementations of this algorithm for large $ N_{\mathrm{L}} $.

The corresponding single ion phase gates in this step are also described by Eq.\eqref{phasegate}
with phases
\[ \theta_j = - 2 \pi \cdot 2^{-(k-j)}
\]
for $ j=1,2,...,k-1 $ and
\[ \theta_k = - 2 \pi \sum_{m=1}^{k-1} 2^{-(k-m)}
\]
for the $k$-th ion.
These coefficients are chosen such that they give the desired controlled phase gates and also discard undesired interactions among the logic ions.

\subsection{D.~~~Exact and approximated gate dynamics}

\paragraph*{Exact solutions---}
Exploiting the possible time dependence of $\Omega_{\alpha i}(t)$ leads to exact solutions of the desired dynamics.
The Rabi frequencies required for this can be obtained by considering the singular value decomposition for the r.h.s. of Eq.~\eqref{eq:cond1}. Let $s_n>0$ be the corresponding singular values (of dimension Hz) and $v^{n}\in\mathds{R}^{\Nc}$ and $w^n\in\mathds{R}^{\Nl}$ the left and right singular vectors, respectively. Eq.~\eqref{eq:cond1} then is equivalent to
\begin{equation}
  \int_0^\tau\!\mathrm{d}t\, \Omega_{\C i}(t)\Omega_{\L j}(t) = \sum_{n=1}^{\Nl} s_{n} v_i^{n} w_j^n \label{eq:SVD}.
\end{equation}
Let $P_n(t)$ be any set of orthonormal functions on $ \left[0, \tau \right] $, i.e.
\begin{align*}
\int_0^\tau\!\mathrm{d}t\, \mathrm{P}_n (t) \mathrm{P}_m (t) = \delta_{nm}.
\end{align*}
When the Rabi frequencies are chosen as
\begin{align*}
\Omega_{\C i}(t) &= \sum_{n=1}^{\Nl} s_{n} \dfrac{1}{r_n} v_i^{n} \, \mathrm{P}_n (t) \\
\Omega_{\L j}(t) &= \sum_{m=1}^{\Nl} r_m \, w_j^{m} \, \mathrm{P}_m (t)
\end{align*}
with $ r_n \in \mathds{R} $ Eq.~\eqref{eq:SVD} is fulfilled. Possible choices for $ \mathrm{P}_n(t) $ include piecewise constant functions (in the spirit of segmented MS gates explored in \cite{Choi2014}) or e.g. Legendre polynomials. A change in the sign of Rabi frequencies may be necessary therein and can be reached by a change of $ \pi $ in the laser phase on the respective ion. The additional vector $r$ can be used to further adapt the required Rabi frequencies.

\paragraph*{Approximation scheme---}
Alternatively, and much more appropriate for the context of optical ion clocks, an approximate solution to condition~\eqref{eq:cond1} can be achieved with time-independent Rabi frequencies: Here the left hand side of \eqref{eq:cond1} becomes $\Gamma^{\C\L}_{ij}= \Omega_{\C i}\Omega_{\L j}\tau$ which is a matrix of rank one. The r.h.s. of Eq.~\eqref{eq:cond1}, which generically has full rank ($\Nl$), can be approximated by a rank one matrix as follows: If the detuning $ \delta_1 \ll \Delta \nu = \nu_1- \nu_2 $ is small as compared to the mode-spacing the MS interaction is mostly mediated via the highest lying collective mode with small corrections from other modes.

When the contribution of lower lying modes is neglected the interaction factorises as $ 1/\Delta^{\alpha\beta}_{ij} \approx \eta_{\alpha i} \eta_{\beta j}/\delta_1 $ and also becomes a rank one matrix (\textit{single mode approximation}).
Therefore the Rabi frequencies $ \Omega_{\C i}= \delta_1/4\eta_{\C i}^1 $ and $ \Omega_{\L j}= 2^{-(j+1)} \> \delta_1/\eta_{\L j}^1 $ with $ i=1,\cdots, \Nc $ and $ j=1,\cdots, \Nl $ give an approximate solution.

A better result can be achieved by finding the singular value decompostion of the r.h.s. of Eq.~\eqref{eq:cond1}, and setting all but the largest singular values to zero (\textit{rank one approxiamtion}), such that Eq.~\eqref{eq:cond1} becomes
\begin{equation}
  \Omega_{\C i}\Omega_{\L j}\tau = s_{1} v_i^{1} w_j^1.
\end{equation}
This is solved by setting
\begin{align*}
  \Omega_{\C i}&=\sqrt{\frac{s_{1}}{\tau}}\,\frac{1}{r} v_i^1 &
  \Omega_{\L i}&=\sqrt{\frac{s_{1}}{\tau}}\,r w_i^1
\end{align*}
for any convenient choice of $r\in\mathds{R}$. In the case study discussed in the main text $r$ was chosen such that the largest Rabi frequency $\Omega_{\L 1}$ still has the same values as in the single mode approximation, that is $\Omega_{\L 1}=\delta_1/4\eta^1_{\L1}$. This ensures that the adiabaticity condition in the MS gates is fulfilled for all ions.

\end{document}